# Improvement of Control System Performance by Modification of Time Delay

Salem Alkhalaf
Computer Department, College of Science and Arts
Qassim University
Alrass City, Qassim, KSA

*Abstract*—This paper presents a mathematical approach for improving the performance of a control system by modifying the time delay at certain operating conditions. This approach converts a continuous time loop into a discrete time loop. The formula derived is applied successfully to an applicable control system. The results show that the proposed approach efficiently improves the control system performance. The relation between the sampling time and the time delay is obtained. Two different operating conditions are examined to assess the proposed approach in improving the performance of the control system.

## I. INTRODUCTION

A distributed control system (DCS) has many interconnected devices, which exchange data through the communication networks, such as home automation, factories, space shuttles, and industrial control Ethernet. The performance of the DCS network is assessed according to the ability of the network and its communication link to transmit the signal (bits) through the network with minimum delay and distortion. A delay in the process should be considered when designing the DCS network. The control loop performance over a network control system has been investigated and studied [1][2][3][4]. The real-time system, control system, and communication system have been studied. The performance of the control loop DCS network depends on many factors such as communication protocols, reducing the communication with dead bands, sampling time, and scan time. The time delay is considered when a design methodology for optimizing the performance of the distributed control system is presented [5]. Compensation for delay time uncertainties on industrial control Ethernet network has been investigated to prove how important delay time is in the control system performance [6]. The control system delay is the summation of the sensor to controller delay, controller calculation delay, and controller to actuator delay.

An adaptive sampling scheme has been presented [7] to ensure that the control delay is less than the sampling period in the steady state and uses the maximum tolerable delay at a specified sampling period to ensure stable transformation from one sampling interval to another.

Echo state neural networks have been used to improve the shape recognition performance of the sendzimir mill control system [8]. Modification of the control system based on artificial intelligence has been used successfully to improve the performance of existing coal-fired thermal power plants. A parameter prediction model based on an artificial neural network has been used to analyze the effect of advanced control on the combustion process, which lead to the development of a self-learning controller [9].

An adaptive robust control law for linear systems with norm-hounded parameter uncertainties has been developed for a robust control system. Online information of the actual system is used to tune the interpolation coefficient. This control scheme overcomes the shortage in a linear robust control with fixed parameters [10].

The neuro-fuzzy approach has been applied to the delay compensator to reduce variable sampling to actuation delays effect in the distributed control system. The approach proposed adding a compensator to an existing distributed system to overcome the degradation of the control performance that results from the variable sampling to actuation delay [11].

An improvement of existing coal fired thermal power plants performance by control systems modifications is discussed. The such system is applied via implementation of advanced combustion control concepts in selected Western Balkan thermal power plant, and particularly those based on artificial intelligence as part of primary measures for nitrogen oxide reduction in order to optimize combustion and to increase plant efficiency. Advanced self-learning controller has been developed and the effects of advanced control concept on combustion process have been analyzed utilizing artificial neural-network based parameter prediction model [12]. A discrete-time control systems performance has been optimized based on network-induced delay. The technique solving optimal tracking problem for single-input single-output (SISO) linear time-invariant discrete-time systems over communication channel with network-induced delay in the feedback path [13]. The output feedback control problem of an interconnected time-delay systems with prescribed performance has been solved and investigated. To obtain such valid solution, a few of the existing results consider the prescribed performance control in the nonlinear interconnected time-delay systems [14].

Keeping the sensor data validity while exercising timely control is crucial in real-time sensing and control systems. The objective of scheduling algorithms deployed in such systems is to keep the validity of the real –time sensor data. This approach leads to maximize the schedule ability of update transactions with minimum update workload, hence the control system performance is adapted to be active on time [15]. Robust iterative learning control design for uncertain time-delay systems based on a performance index has been investigated, analyzed and discussed [16]. A robust iterative learning control (ILC) for uncertain time-delay systems has been designed based on a performance index for the error system. The Lyapunov-like approach can be applied to design robust ILC for uncertain systems with time-varying delay or multiple time delays. Enhancing the performance bounds of the multivariable control systems have high degree of thinness have been discussed, analyzed, and investigated [17]. A real-time implementation of fault-tolerant control (FTC) systems with performance optimization have been discussed, investigated and analyzed [18].

This paper aims to present a mathematical approach for improving the performance of a control system by converting the system continuous loop into a discrete loop. The formula derived is applied successfully to the studied system. The results show that the sampling time and the delay time are optimized. The sampling time against the time delay is an interested point.

## II. CONTROL SYSTEM MODEL

The control system is shown in Figure 1 [19]. The simplest model is given by discrete-time control models obtained from continuous time models that include a constant time delay in the mathematical formulation. The continuous time state space model of the linear time invariant system can be described by the following standard form [20][21]:

$$dx(t)/dt = Ax(t) + Bu(t) \quad (1)$$

$$y(t) = cx(t) + Du(t) \quad (2)$$

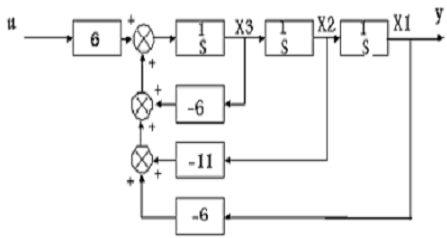

Fig. 1. Control System Block Diagram

Applying equations 1 and 2 to the control system, the system equations can be obtained in the matrix form:

$$\begin{bmatrix} X'1 \\ X'2 \\ X'3 \end{bmatrix} = \begin{bmatrix} 0 & 1 & 0 \\ 0 & 0 & 1 \\ -6 & -11 & -6 \end{bmatrix} \begin{bmatrix} X1 \\ X2 \\ X3 \end{bmatrix} + \begin{bmatrix} 0 \\ 0 \\ 6 \end{bmatrix} u \quad (3)$$

And the output equation is:

$$y = \begin{bmatrix} 1 & 0 & 0 \end{bmatrix} \begin{bmatrix} x1 \\ x2 \\ x3 \end{bmatrix} \quad (4)$$

## III. PREPARE DELAY TIME AND SAMPLING TIME APPROACH

The classical model for discrete-time control systems assumes that the control algorithm is executed instantaneously at every sampling period h. Consequently, equidistant sampling and actuation are assumed.

Based on these assumptions, for periodic sampling with constant period h, the discrete time system can be described by:

$$x(kh+h) = x\,kh\,e^{Ah} + \frac{B}{A}[e^{A(h-\tau)} - 1]u\,kh + \frac{B}{A}e^{Ah}(1 - e^{-A\tau})u(kh - h) \quad (5)$$

Where:

h: Sampling time,

k: Number of control loop execution,

τ: Delay time,

A: Feedback transmission factor.

Multiplying equation 5 by B/A gives:

$$x\frac{A}{B}[kh + h - kh\,e^{Ah}] + ukh + e^{Ah}uh = e^{-A\tau}e^{Ah}hu \quad (6)$$

Reducing the previous equation gives:

$$e^{-A\tau} = \frac{A}{B}e^{-Ah}\frac{x}{u}(k - ke^{Ah} + 1) + k\,e^{-Ah} + 1 \quad (7)$$

Assume:

$$\frac{x}{u} = y$$

$$e^{-A\tau} = \frac{A}{B}e^{-Ah}(k - ke^{Ah} + 1)\,y + k\,e^{-Ah} + 1 \quad (8)$$

$$\tau = -\frac{1}{A}\ln[\frac{A}{B}e^{-Ah}(k - ke^{Ah} + 1)y + ke^{-Ah} + 1] \quad (9)$$

A is always –ve (negative feedback)

Thus, the equation above is valid if and only if

$$\frac{A}{B}e^{-Ah}(k - ke^{Ah} + 1)y + ke^{-ah} > 0 \quad (10)$$

or

$$\frac{A}{B}(k - ke^{Ah} + 1)y + k > 0 \quad (11)$$

i.e.,

$$y < \frac{B}{A} \frac{-k}{(1+k-ke^{Ah})} \quad (12)$$

## IV. SIMULATION RESULTS

*Case I*

From the loop of Figure 1 and the substitution in equation 12 by the given constant A=–6, B=6, k=32, h=0.001…………0.005 sec, the results are shown in Table 1.

TABLE I. SAMPLING TIME AGAINST DELAY TIME

| h | y | τ |
|---|---|---|
| 0.001 | 26 | 0.118 |
| 0.002 | 23 | 0.034 |
| 0.003 | 20 | 0.078 |
| 0.004 | 18 | 0.05 |
| 0.005 | 16 | 0.107 |

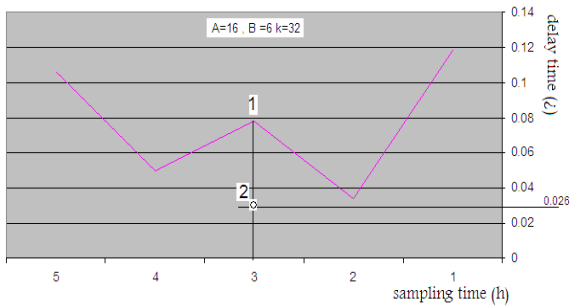

Fig. 2. A=–6 Primary Rough Relation Before Correction

Figure 2 depicts the primary rough relation between the control system delay time and the sampling time. Point 1 is irregular so piece wising of the curve will give a new point as (0,003,0.026). The table after correction is shown in Figure 2.

TABLE II. SAMPLING TIME AGAINST DELAY TIME AFTER CORRECTION

| h | y | τ |
|---|---|---|
| 0.001 | 26 | 0.118 |
| 0.002 | 23 | 0.034 |
| 0.003 | 20 | 0.026 |
| 0.004 | 18 | 0.050 |
| 0.005 | 16 | 0.108 |

The least square root method is used for the results after correction, which are given in Table 2.

Related to the quadratic equation, y=a+bx+cx2, where a, b, and c are constants, normal equations are:

$$N a + b\sum h_i + c\sum h_i^2 = \sum \tau_i \quad (13)$$

N=number of points (h's).

Substituting the values of h, n, and τ in equation 13, a, b, and c are obtained:

a=0.23, b=–136, c=22571

These values are substituted in equation 13.

$$\tau = 0.23 - 136\,h + 2257\,h^2 \quad (14)$$

Differentiate the previous equation:

$$\frac{d\tau}{dh} = 0 = -136 + 45142h$$

h=0.003, τ =0.025

Thus, the lowest point is (0.003,0.025).

Taking the different values of h and substituting them in equation 14 give the results shown in Table 3.

TABLE III. SAMPLING TIME AGAINST DELAY TIME

| h | y | τ |
|---|---|---|
| 0.001 | 26 | 0.118 |
| 0.002 | 23 | 0.034 |
| 0.003 | 20 | 0.026 |
| 0.004 | 18 | 0.050 |
| 0.005 | 16 | 0.108 |

| h | τ |
|---|---|
| 0.001 | 0.117 |
| 0.002 | 0.048 |
| 0.003 | 0.025 |
| 0.005 | 0.114 |

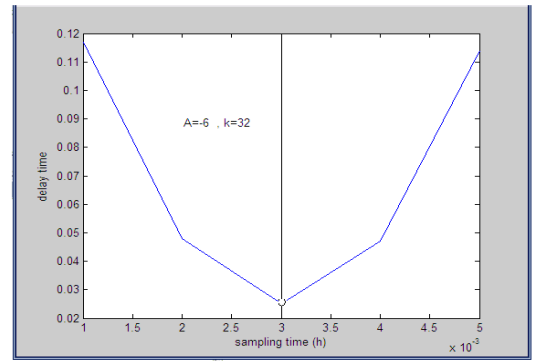

Fig. 3. Final Sampling Time Against Delay Time (A=–6, k=32)

Figure 3 depicts the primary rough relation between the control system delay time and the sampling time after correction. Comparing Figure 2 and Figure 3 shows the efficiency of the proposed approach in improving the control system performance to obtain an optimized point.

*Case 2*

The second operating point is considered A=–11, B =6, k=32, h=0.001,0.002,…,0.005 sec. The results are shown in Table 4.

TABLE IV. SAMPLING TIME AGAINST THE DELAY TIME

| h | y | τ |
|---|---|---|
| 0.001 | 12 | 0.109 |
| 0.002 | 10 | 0.059 |
| 0.003 | 8.5 | 0.019 |
| 0.004 | 7 | 0.085 |
| 0.005 | 6 | 0.108 |

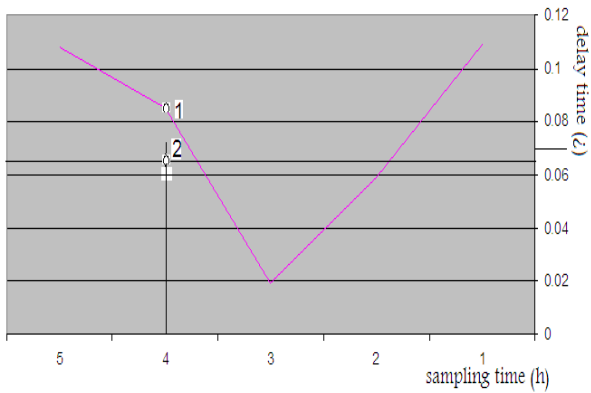

Fig. 4. A=−11, k=32 Primary Rough Relation Before Correction

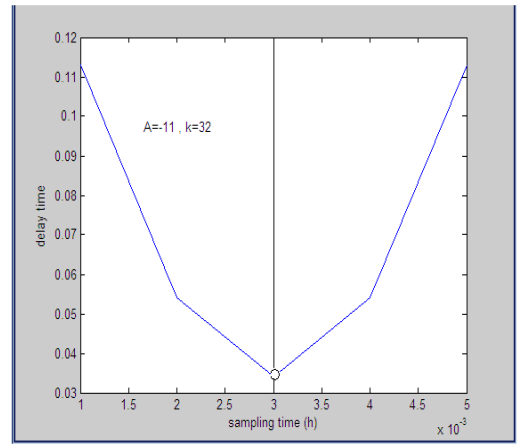

Fig. 5. Final Sampling Time Againstthe Delay Time (A=−11, k=32)

In Figure 4, the relation should be piece wised to give the results shown in Table 5

TABLE V. SAMPLING TIME AGAINST THE DELAY TIME

| h, sec. | y | $\tau$, sec. |
|---|---|---|
| 0.001 | 12 | 0.109 |
| 0.002 | 10 | 0.059 |
| 0.003 | 8.5 | 0.019 |
| 0.004 | 7 | 0.063 |
| 0.005 | 6 | 0.108 |

The least square root method is used as in the first case, so that the constant value is calculated.

a=0.21, b=−117.23, c=19571

i.e., at A=−11, k=32, the quadratic equation is

$$0.21 - 117.23\,h + 19571\,h^2 = 0 \quad (13)$$

Differentiate the previous equation

$$\frac{d\tau}{dh} = -117.23 + 3914h = 0$$

Thus, h=0.003, $\tau$ =0.034
The lowest point is (0.003, 0.034).

These values are substituted in equation 13, and the corrected delay time is shown in Table 6.

TABLE VI. SAMPLING TIME AGAINSTTHE DELAY TIME

| h sec | sec$\tau$ |
|---|---|
| 0.001 | 0.113 |
| 0.002 | 0.054 |
| 0.003 | 0.034 |
| 0.004 | 0.054 |
| 0.005 | 0.113 |

Figure 5 depicts the relation between the corrected sampling time and the delay time after the optimized point according to h=3 m sec.

Comparing Figure 4 and Figure 5 shows that the system performance after correction is improved, and an optimization point corresponds to the sampling time, at h=0.003 sec.

## V. CONCLUSION

In this paper, the following points can be concluded: (1) The sampling time and the delay time are optimized to improve the performance of a control system. (2) Two different operating points are considered to assess the efficiency of the proposed approach in improving the control system performance. (3) The point of the minimum delay time should be studied to assess the system performance. (4) When the sampling time becomes close to zero, the performance will be close to that of the continuous system. (5) The equation proves that when the sampling time increased the delay time increased, and vice versa.